\documentstyle[aps,preprint,epsf,graphics]{revtex}
\begin{document}

\Large

\center{\bf AdS/CFT Correspondence and QCD with}
\center{\bf Quarks in Fundamental Representations}

\normalsize

\center{Paul H. Frampton }

{\it Department of Physics and Astronomy,}

{\it University of North Carolina,Chapel Hill, NC 27599, USA.}

\bigskip
\bigskip
\bigskip
\bigskip
\bigskip
\bigskip
\bigskip
\bigskip
\bigskip
\bigskip
\bigskip
\bigskip

\abstract
The most straightforward use of AdS/CFT correspondence gives versions of QCD
where quarks are in adjoint representations. Using an asymmetric orbifold
approach we obtain nonsupersymmetric QCD
with four quark flavors in fundamental representations of color.~~~~~~~~~~~~~~~

\newpage

\bigskip
\bigskip

\newpage

The interplay between gauge field theories and string theory is one
of the most fertile in high-energy theory. Indeed string theory had
its beginnings as an attempted theory of strong interactions
\cite{V,DRM} to be replaced by the gauge field theory of
quantum chromodynamics (QCD)\cite{N,FGL}. Nevertheless the
interconnection between these theories constantly portrays them
not as competitors but most recently as dual descriptions. The
notion that strings describe some large $N_{colors}$ limit
of QCD was suggested already by 't Hooft\cite{largeN} in 1974.
An excellent review is provided by the 1987 book by Polyakov\cite{Polyakov}.
A large step forward was taken with the identification
by Maldacena\cite{Maldacena} of the AdS/CFT correspondence
For example, compactifying a superstring on an
$AdS_5 \times S^5$ manifold there is a duality of descriptions
of either a type IIB superstring in ten spacetime dimensions or
an ${\cal N} = 4$ supersymmetric $SU(N)$ gauge field theory in
four spacetime dimensions.

\bigskip

This correspondence provides a powerful tool to
investigate gauge field theory. Independently of
whether the superstring can provide a correct theory
of quantum gravity the AdS/CFT correspondence can
suggest interesting models for non-gravitational 
physics. Beyond being merely a tool, 
it can suggest directions to extend the
standard model and even additional
particles that may
be lying in the TeV regime awaiting
discovery\cite{FRT}. As a tool, it may be used to study QCD and
very interesting results have been obtained 
by Polchinski and Strassler \cite{PS,PS2,BT} about the
relationship of QCD to string theory from 
the AdS/CFT correspondence. This is of special interest
in view of the above history because the string theory was 
originally displaced by QCD thirty years ago.

\bigskip

This recent work on string QCD sheds light on how QCD describes 
hard scattering by pointlike constituents\cite{Brodsky} whereas
the string gave only infinitely soft
scattering. The resolution come from the importance
of the curved geometry of the fifth dimension
in the $AdS_5$ manifold.

\bigskip

Here we make an observation about the group theory
of orbifolding $AdS_5 \times S^5$ and the avoidance
of quarks in adjoint representations. In real QCD
with gauge group $SU(3)$ the quarks come in six flavors
$(u, d, s, c, b, t)$ which transform as fundamental
triplets, not adjoint octets, of the color $SU(3)$
gauge group.

\bigskip

It is indeed much easier to obtain adjoint quarks than fundamental 
quarks from AdS/CFT derivations. For example, with the manifold
$AdS_5 \times S^5$ the GFT is an ${\cal N}=4$
supersymmetric $SU(N)$ gauge field theory in which the
fermions are gauginos in the adjoint representation.
This is sometimes called ${\cal N}=4$ QCD and has 
considerable similarity with QCD. The two
principal differences are
the presence of supersymmetry and the fact that the quarks 
are in adjoint representations rather than
in fundamental representations.

\bigskip

The reconciliation of these two
differences with QCD can be arranged
simultaneously by the following asymmetric orbifold procedure.

\bigskip

To break supersymmetry from ${\cal N} = 4$
to ${\cal N} = 0$
we replace the manifold $AdS_5 \times S^5$ by
the orbifold
$AdS_5 \times S^5/\Gamma$ where $\Gamma$ is a finite subgroup
of the $SU(4)$ isometry
of $S^5$ and such that
$\Gamma \not\subset SU(3) \subset SU(4)$
for any choice of the $SU(3)$ subgroup.
This ensures that ${\cal N} = 0$
or no supersymmetry in the gauge field theory.
There is no advantage in using non-abelian $\Gamma$
so we choose abelian $\Gamma = Z_p$
which leads to a semi-simple gauge group
$SU(N)^p$.

\bigskip

QCD has only non-chiral quarks so we need to choose an
embedding with ${\bf 4} \equiv {\bf 4^*}$ of $SU(4)$.
Defining $\alpha^p = 1$ or $\alpha = \exp (2 \pi i /p)$ the embedding is defined by
${\bf 4} = (\alpha^{A_1}, \alpha^{A_2}, \alpha^{A_3}, \alpha^{A_4})$
and
for ${\cal N} = 0$
one requires all $A_{\mu}$ ($\mu = 1, 2, 3, 4)$ to be non-zero
and that the sum 
$\sum_{\mu=1}^{\mu=4} A_{\mu} = 0$ (mod $p$).
The fermions which survive the orbifolding
are those which are invariant
under a product of a $Z_p$
transformation and a $SU(N)^p$
gauge transformation.
These fermions are in the bifundamental representations
of $SU(N)^p$:

\begin{equation}
\sum_{i=1}^{i=p} \sum_{\mu=1}^{\mu=4} (N_i, \bar{N}_{i + A_{\mu}})
\label{chiralfermions}
\end{equation}

\bigskip

The complex scalars which survive are also those which are
invariant under a product of a 
$Z_p$ transformation and a $SU(N)^p$ gauge transformation but now follows from
the real ${\bf 6} \equiv (v_i, v_i*)$ of $SU(4)$
with $i = (1, 2, 3)$
and 
$v_i = (\alpha^{a_1}, \alpha^{a_2}, \alpha^{a_3})$.
Here $a_1 = (A_2 + A_3)$,
$a_2 = (A_3 + A_1)$ and $a_3 = (A_1 + A_2)$ although we note
that the subscript orderings of $A_{\mu}$, $a_i$
are arbitrary and we can therefore
choose $A_1 \le A_2 \le A_3 \le A_4$
as well as $a_1 \le a_2 \le a_3$.
The complex scalars now transform under
$SU(N)^p$
as

\begin{equation}
\sum_{j=1}^{j=p} \sum_{i=1}^{i=3} (N_j, \bar{N}_{j \pm a_1})
\label{complexscalars}
\end{equation}

\bigskip

Both Eq.(\ref{chiralfermions}) and Eq.(\ref{complexscalars})
can be conveniently displayed in quiver or moose diagrams
as we shall illustrate for $p=3$ in Figure 1. ~~~~~~~~~

\bigskip

For simplicity we choose the lowest
possible value of $p$ ($p=3$) which works.

\bigskip

Note that for $p=2$ the only choice for $A_{\mu}$ is
$A_{\mu}=(1, 1, 1, 1)$ and
consequently $a_i=(0, 0, 0)$ which means the quiver diagram
for scalars, from Eq.(\ref{complexscalars}), is disconnected into two
separate $SU(N)$s and no progress toward QCD is possible.
Choosing $p=3$ is, however, sufficient.
In this case the unique
choice to attain ${\cal N} = 0$
is $A_{\mu} = (1, 1, 2, 2)$
and therefore $a_i = (0, 0, 1)$.
The scalar quiver is shown in Figure 1(a).
Each $SU(N)$ has two adjoints and two $(N + N^*)$
fundamentals of scalars.

\bigskip

If we now spontaneously break the
$SU(N)^p$ symmetry in a way
which respects the $Z_p$ symmetries of the quiver diagram,
and identify the QCD gauge group
as the diagonal subgroup
of the three $SU(N)$ groups
then the quarks will again be in adjoints
just as in the ${\cal N} = 4$ case
but without the supersymmetry.

\bigskip

Instead we choose an asymmetrical symmetry breaking
where two of the $SU(N)$s, say the two
lower nodes in Figure 1(a) are completely
broken which is straightforward using
VEVs of the available scalars in an asymmetric
potential.

\bigskip

The third node, say the upper vertex of the triangular quiver in
Figure 1(a), is identified with
the gauge group
of QCD. There are two color adjoints and two fundamentals of scalars
which can be made massive.

\bigskip

Finally, the chiral fermions
are given by Eq. (\ref{chiralfermions}) and
are depicted in the quiver diagram of Figure 1(b).
With respect to the unbroken QCD $SU(N)$ the fermions are non-chiral and occur in four
flavors of $(N + N^*)$.
No adjoint quarks appear.  ~~~~~~~~~~~~~~~~~~~~~~~~~~~

\bigskip

Four flavors occur because of the {\bf 4} of the $SU(4)$
isotropy of $S^5$. Increasing to $p \ge 4$ does not increase
the number of flavors and has no other advantage so the
$p = 3$ case of Figure 1 is the
simplest AdS/CFT model for nonsupersymmetric
QCD with quarks in fundamental representations.

\bigskip

The dynamical studies in \cite{PS,PS2}
presumably carry over to the present case
since they depend only on the AdS geometry. Using the 
present AdS/CFT version of string QCD 
will, however,  help arrange the correct 
color group theory factors to appear in the calculations.

\bigskip
\bigskip
\bigskip

\newpage

This issue arose in a seminar by Professor Stanley J. Brodsky
in Spring 2003.  This work was supported in part by the
Office of High Energy, US Department
of Energy under Grant No. DE-FG02-97ER41036.

\newpage

\begin{figure}
\begin{center}
\epsfxsize=2.5in
\ \epsfbox{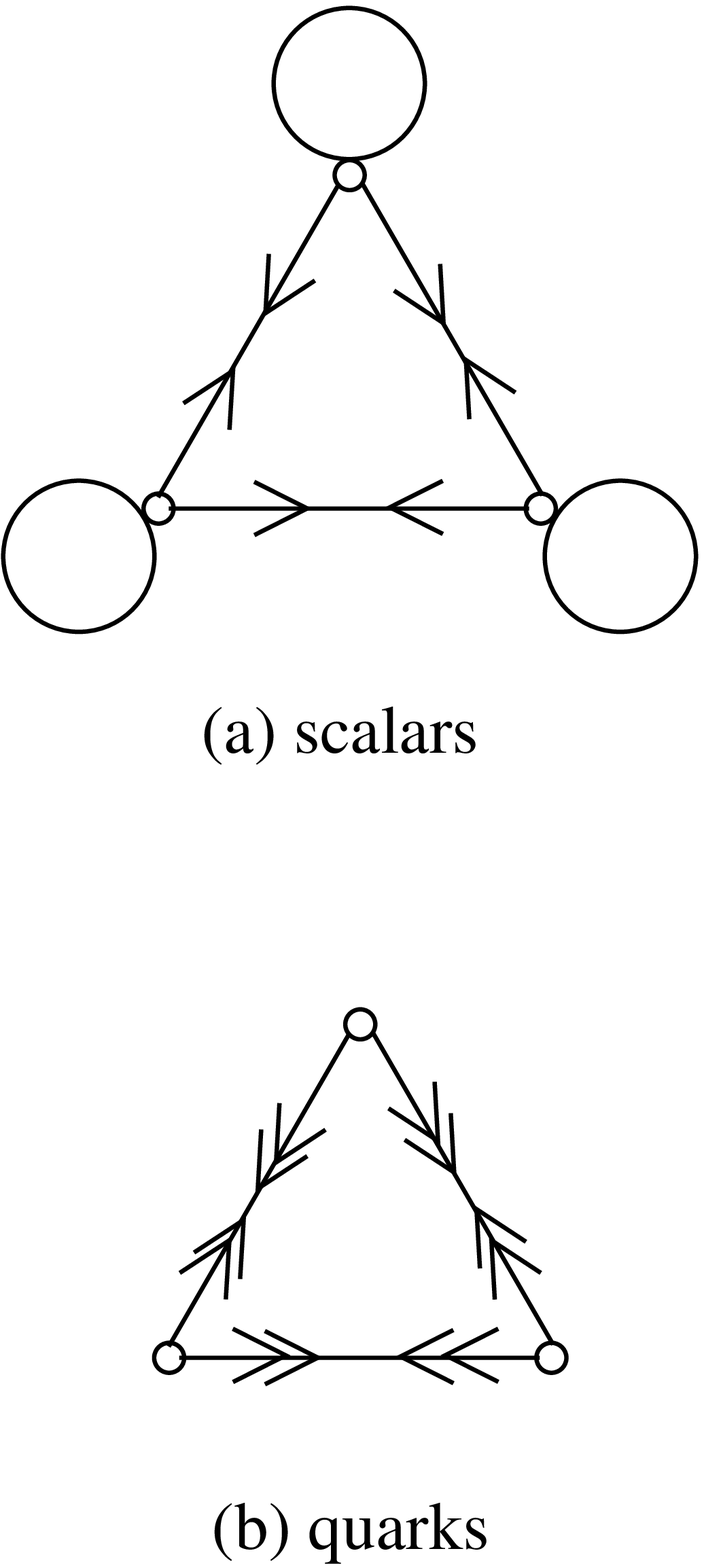}
\end{center}
\end{figure}

\bigskip
\bigskip
\bigskip

\LARGE

\begin{center}

Figure 1.  Quiver Diagrams

\end{center}

\end{document}